\documentclass[prl,twocolumn]{revtex4-1}
\usepackage{graphicx}
\usepackage{graphpap}
\usepackage{graphics}
\usepackage{amssymb}
\usepackage{amsmath}
\usepackage{enumitem}
\usepackage{color}
\usepackage{xfrac}
\usepackage{multirow}
\usepackage{array}

\begin{document}

\title{Tunable mechanical coupling between driven microelectromechanical resonators}

\author{G.~J.~Verbiest$^{1}$}
\author{D.~Xu$^{1}$}
\author{M.~Goldsche$^{1,2}$}
\author{T.~Khodkov$^{1,2}$}
\author{S.~Barzanjeh$^{3,4}$}
\author{N.~von~den~Driesch$^{2}$}
\author{D.~Buca$^{2}$}
\author{C.~Stampfer$^{1,2}$}
\affiliation{$^{1}$JARA-FIT and 2nd Institute of Physics, RWTH Aachen University, 52074 Aachen, Germany}
\affiliation{$^{2}$Peter Gr\"{u}nberg Institute (PGI-8/9), Forschungszentrum J\"{u}lich, 52425 J\"{u}lich, Germany}
\affiliation{$^{3}$Institute for Quantum Information, RWTH Aachen University, 52056 Aachen, Germany}
\affiliation{$^{4}$Institute of Science and Technology Austria, 3400 Klosterneuburg, Austria}

\begin{abstract}
We present a microelectromechanical system, in which a silicon beam is attached to a comb-drive actuator, that is used to tune the tension in the silicon beam, and thus its resonance frequency. By measuring the resonance frequencies of the system, we show that the comb-drive actuator and the silicon beam behave as two strongly coupled resonators. Interestingly, the effective coupling rate ($\sim 1.5$ MHz) is tunable with the comb-drive actuator ($+10$\%) as well as with a side-gate ($-10$\%) placed close to the silicon beam. In contrast, the effective spring constant of the system is insensitive to either of them and changes only by $\pm 0.5\%$. Finally, we show that the comb-drive actuator can be used to switch between different coupling rates with a frequency of at least 10 kHz.
\end{abstract}

\maketitle

% introduction
%\section{Introduction}

Micro- and nanoelectromechanical resonators have attracted much attention thanks to their potential application as sensors \cite{srivastava2004guest,mori1999fast,tabak2010mems,disseldorp2010mems,martin2013graphene,schwab2005putting,wallquist2009hybrid,eisert2004towards}, filters~\cite{yang2001cmos,gouttenoire2010digital,piekarski2001surface}, amplifiers~\cite{rebeiz2004rf,karabalin2011signal}, and logic gates~\cite{tsai2008design}, with frequencies varying from the kHz to the GHz range~\cite{cleland1996fabrication,nguyen1998mixing,ekinci2002balanced}.
By applying tension, the resonance frequencies can be tuned in-situ ~\cite{buks2002electrically,zhu2009strategies,challa2008vibration}.
This is usually done by applying a DC voltage to a nearby gate to induce a deformation of the resonator via the electrostatic potential~\cite{lei2015mechanical,no2002single,piazza2004voltage}. However, one does not only induce tension by applying an electrostatic potential, but one also changes the number of charge carriers~\cite{lassagne2008ultrasensitive,grogg2008multi,sun2007physics}, which both affect the properties of the resonator~\cite{li2003thermal,lifshitz2000thermoelastic}.
An ideal candidate to circumvent this problem is the comb-drive actuator, which is nowadays widely used to detect accelerations and rotations as well as to manipulate, stretch, or move objects~\cite{tang1990electrostatic,lin1992microelectromechanical,su2005resonant,lee1992polysilicon,garcia1995surface}. By mechanically fixing a resonator to a comb-drive, one can purely mechanically induce strain in the resonator.
Depending on the type of comb-drive actuator, this induced strain can be either tensile or compressive. The former allows the disentanglement of strain and charge carrier density effects in mechanical resonators, whereas the latter is particularly relevant to gain control over buckling modes in mechanical resonators~\cite{saif2000tunable,sulfridge2004nonlinear}.

% FIG 1: Device and Setup
\begin{figure}[!h]
	\begin{center}
		\includegraphics[trim=0 0 0 -2, width=78mm]{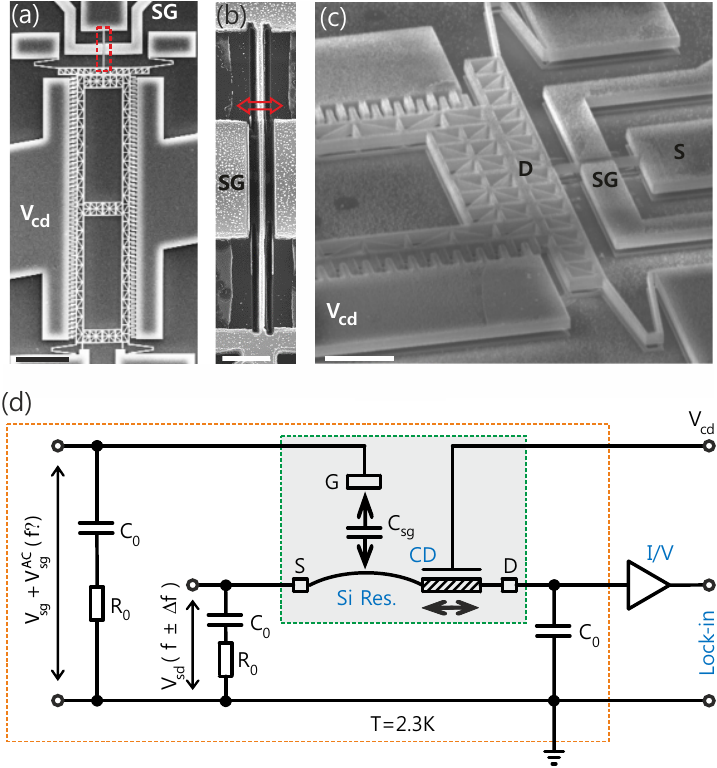}
	\end{center}
		\setlength{\abovecaptionskip}{-0.1cm}
	\caption{
		Scanning electron microscope (SEM) images of (a) the comb-drive actuator and (b) a zoom in of the dashed red box in panel (a) as well as (c) a tilted angle view of the device. The scale bars indicate 20~$\mu$m, 2~$\mu$m and 10~$\mu$m, from left to right. The electronic circuit depicted in (d) is used to measure the resonance frequency, where the inner green box indicates the device, and the orange box indicates the printed circuit board (PCB) in the cryostat at $T=2.3$ K. An AC signal ($V_{sd}(f\pm\Delta f)$) is used to drive the current from source (S) to drain (D) contacts while a sum of DC and AC voltages ($V_{sg} +V_{sg}^{AC}(f)$) are applied to the side-gate (SG). The down-mixed current at frequency $\Delta f$ through the resonator is measured by an I/V-converter and a lock-in amplifier. The comb-drive actuator is driven by a DC voltage $V_{cd}$. The capacitors $C_0 = 100$ nF and resistors $R_0 = 50$ $\Omega$ are chosen for impedance matching and decoupling any high frequency signals.}
	\label{fig1}
\end{figure}

% FIG 2: Raw data
\begin{figure*}[!hbt]
	\begin{center}
		\includegraphics[width=\linewidth]{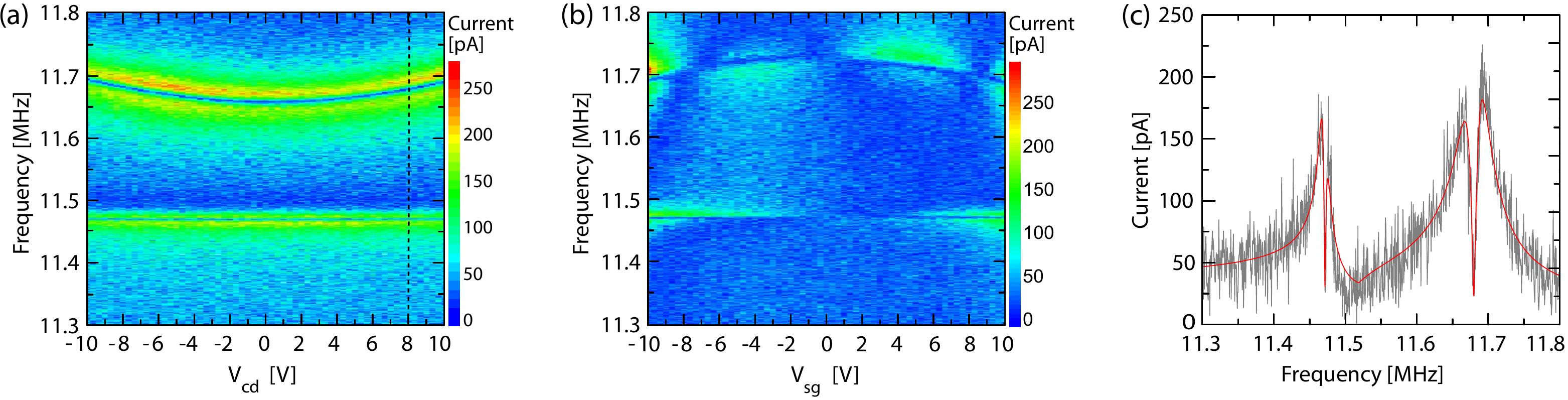}
	\end{center}
	\setlength{\abovecaptionskip}{-0.3cm}
	\setlength{\belowcaptionskip}{-0.3cm}
	\caption{Measured down-mixed current as functions of (a) the applied $V_{cd} $ at $V_{sg} = 10$ V and (b)  $V_{sg}$ at $V_{cd}  = 10$ V. The data of the vertically dashed black line in (a) are shown in (c). The red line is the fit to Eq.~\ref{Eq-Current}.}
	\label{fig2}
\end{figure*}

Coupled micro- and nanomechanical resonators are known to show sensitive sympathetic oscillation dynamics that show better performance in potential applications than a single resonator~\cite{spletzer2006ultrasensitive,okamoto2011high,shim2007synchronized,karabalin2009nonlinear,faust2012nonadiabatic}. One major challenge for this type of resonators is an in-situ control over the coupling~\cite{masmanidis2007multifunctional}. From a fundamental point of view, a tunable mechanical coupling is interesting because of the feasibility to coherently manipulate phonon populations in coupled resonators~\cite{okamoto2013coherent,PhysRevA.93.033846} and to create intrinsically localized modes~\cite{sato2003observation}. From an applied point of view, researchers are able to build novel mass sensors, bandpass filters, and single-electron detectors ~\cite{spletzer2008highly,okamoto2009optical,pourkamali2005electrically}. 
Fast in-situ control over the coupling between mechanical resonators is an important step towards mechanical logic gates~\cite{mahboob2008bit}. 
Comb drives are suspended microelectromechanical systems based on springs, which typically have a resonance frequency in the (low) kHz range~\cite{patterson2002scanning,legtenberg1996comb,sun2005characterizing}. Once attaching a resonator to this type of micromachine, one could introduce a system of two (strongly) coupled resonators with a potential control of the coupling.
Here, we present an electromechanical system consisting of a comb-drive actuator to control the tensile strain in an integrated silicon beam. We show that the device behaves as a system with two strongly coupled resonators, with a coupling rate $\sim1.5$ MHz, which is tunable up to $+10$\% with the comb-drive actuator as well as down to $-10$\% with a side-gate (SG) placed close to the silicon beam.

%\section{Device and Setup}
The devices are fabricated on a substrate consisting of 500 $\mu$m silicon, 1 $\mu$m SiO$_2$ and 1.6 $\mu$m highly doped silicon. Using standard e-beam lithography, a chromium hard mask is deposited after which reactive ion etching is used to etch away the highly doped silion. After removal of the hard mask, the SiO$_2$ layer is partly etched away with 10\% hydrofluoric (HF) acid solution to suspend the comb-drive actuator and silicon beam. Finally, a critical point dryer (CPD) is used to suspend the structure and protect it from collapsing to the underlying silicon layer. Fig.~\ref{fig1}a shows a typical device with all its electrical contacts of highly doped silicon. The fabricated silicon beam is 13.3 $\mu$m long, 250 nm wide and 1.6 $\mu$m high (Fig.~\ref{fig1}b). A close up is provided in Fig.~\ref{fig1}c showing the high aspect ratio $\sim 0.156$ in our device. 
All the measurements were performed using an amplitude modulated down-mixing technique~\cite{knobel2002single,sazonova2004tunableresonator} in a variable temperature insert at a temperature of $\sim$ 2.3 K.

% FIG 3: Fit results
\begin{figure*}[!ht]
	\begin{center}
		\includegraphics[width=\linewidth]{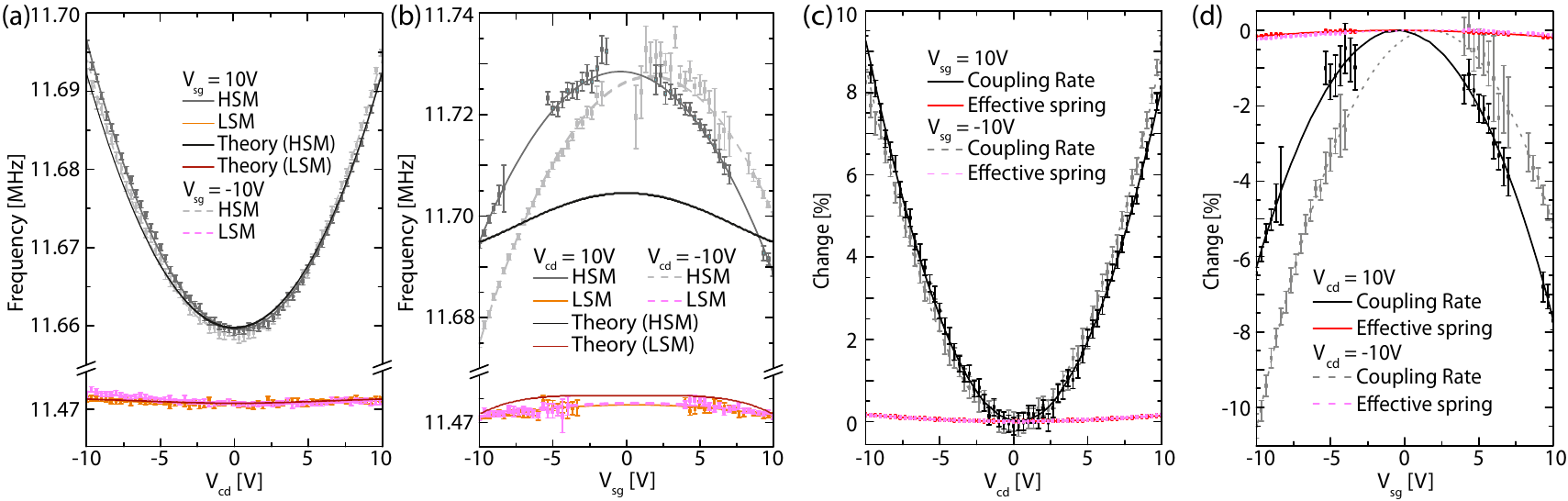}
	\end{center}
	\setlength{\abovecaptionskip}{-0.3cm}
	\setlength{\belowcaptionskip}{-0.3cm}
	\caption{
		The extracted resonance frequencies at $\sim$ 2.3 K as a function of (a) comb-drive voltage $V_{cd}$ ($V_{sg} = \pm 10$ V) and (b) SG voltage $V_{sg}$ ($V_{cd} = \pm 10$ V). The resonance of the HSM corresponds to the silicon beam and the one of the LSM to the comb-drive actuator. The theoretically calculated resonance frequencies are also shown as black and red curves. The system of two coupled resonators is characterized by its effective spring and its coupling rate. (c) and (d) show the relative change in these parameters as a function of $V_{cd}$ and $V_{sg}$ respectively. Interestingly, the change in effective spring is below 0.5\%, whereas the change in coupling can be as high as $\pm10$\%. Moreover, this sign is fully controllable via $V_{cd}$ and $V_{sg}$.}
	\label{fig3}
\end{figure*}

A schematic overview of the electronic measurement circuit is depicted in Fig.~\ref{fig1}d. The capacitors $C_0 =~100$ nF and resistors $R_0 =~50$ $\Omega$ are chosen for impedance matching and for decoupling any high frequency signals. The comb-drive actuator is actuated capacitively by the DC voltage $V_{cd}$ applied between comb fingers. To tune the charge carrier density in the silicon beam, a DC voltage $V_{sg}$ is applied to the SG with a distance 350 nm to the silicon beam, which has a capacitive coupling $C_{sg}$. Both $V_{cd}$ and $V_{sg}$ tune the strain in the silicon beam. In addition, to mechanically excite the silicon beam, an AC voltage $V_{sg}^{AC}$ at frequency $f$ is applied to SG. This leads to a variation in conductance of the silicon beam at the same frequency $f$. An AC voltage $V_{sd}$ at frequency $f-\Delta f$ and, simultaneously, at frequency $f+\Delta f$ is applied to the source (S) - drain (D) contacts of the silicon beam. As a consequence, the current through the silicon beam contains a term at frequency $\Delta f$, which is measured by an I/V-converter and a lock-in amplifier. The current at frequency $\Delta f$ comprises two terms: one stems from the mechanical motion of the beam and the other is due to the variation in charge carrier density $n$. By including an arbitrary phase $\Delta\phi$ shift between these two terms, the current $I(\Delta f)$ is given by~\cite{sazonova2006tunable},

\begin{align}
I(\Delta f) =& \left[ \frac{\frac{\partial C_{sg}}{\partial z} V_{sg}  \frac{z_0}{Q} \cos\left[\beta(f) + \Delta\phi \right]}{\sqrt{\left(1-\left(\tfrac{f}{f_0}\right)^2\right) + \left(\tfrac{f}{f_0 Q}\right)^2}} + C_{sg} V_{sg}^{AC} \right] 2 \frac{\mathrm{d}G}{\mathrm{d}n} V_{sd},\label{Eq-Current}
\end{align}
\noindent
where $\beta(f)=\arctan\left(\tfrac{f_0^2-f^2}{ff_0/Q}\right)$, $f_0$ and $Q$ are the resonance frequency and its corresponding quality factor, $z_0$ is the vibration amplitude of the beam at resonance, $\partial C_{sg}/\partial z$ is the change in capacitive coupling between SG and the silicon beam due to a small beam displacement, and $\mathrm{d}G/\mathrm{d}n$ is the transconductance. Note that the lock-in amplifier measures the absolute current, which can make a single resonance appear as two peaks in $|I(\Delta f)|$, if the phase $\Delta \phi$ is nonzero.
The resonance frequencies of the device are found by sweeping the frequency $f$ while measuring $|I(\Delta f)|$. Typically measured spectra are shown in Fig.~\ref{fig2}a and \ref{fig2}b as a function of $V_{sg} $ at constant $V_{cd}$, and vice versa, respectively, (see Suppl. Mat. II for room temperature measurements). Surprisingly, there are two clear resonant modes.  The low-frequency resonant mode in Figs.~\ref{fig2}a and \ref{fig2}b exhibits almost no tunability as function of $V_{cd}$ or $V_{sg} $, to which we will refer as the ``low sensitive mode'' (LSM), whereas the high-frequency resonant mode is at least $\sim12$ times more sensitive to $V_{cd}$ and $V_{sg} $ and will therefore be called the ``highly sensitive mode'' (HSM).

The resonance frequencies are obtained by fitting Eq.~\ref{Eq-Current} to the raw data (Fig.~\ref{fig2}c). The extracted resonance frequencies in dependence of $V_{cd}$ and $V_{sg}$ are shown in Figs.~\ref{fig3}a and \ref{fig3}b, respectively.
We conclude that tensile strain in the silicon beam induced by the comb-drive actuator via $V_{cd}$ increases the resonance frequency of the HSM, whereas an out-of-plane bending of the silicon beam induced by $V_{sg}$ effectively lowers the resonance frequency. All extracted resonance frequencies of the same measurement lie on a quadratic polynomial, as shown with orange and gray solid curves (and pink and light gray dashed curves) in Figs.~\ref{fig3}a and \ref{fig3}b.
These tunabilities are in agreement with the behavior of the capacitive force between the SG and the silicon beam on the comb-drive actuator, which always scales quadratically with applied voltage, i.e. $V_{cd}^2$ and $V_{sg}^2$, corresponding to the strain hardening and the capacitive softening~\cite{lifshitz2010nonlinear}, respectively. Note that there are slight deviations for the center position at $V_{sg}=0$ (Fig.~\ref{fig3}b) due to residual charges from the fabrictaion process or a capacitve crosstalk between the SG and the comb-drive.

The observed two resonant modes in our system are in good agreement with the theoretical prediction.
In the theoretical model (see Suppl. Mat. I), our device is described by a beam attached to a mass, in which a transverse and an axial drive force act on the beam. The beam is doubly clamped in axial direction and oscillates in the transverse direction, whereas the mass can only move in the axial direction. The resulting dynamics of this system can be expressed by the following equations of motion,
\begin{align}\label{Eqthnbeqomega}
\ddot{v}+\lambda_1v+\varGamma u_l&=0,\nonumber\\
\mu \ddot{u}_l+\lambda_2u_l+\varGamma v&=0,
\end{align}
in which $v$ and $u_l$ denote the time dependent part of the oscillations of the beam and the mass, respectively, i.e. the silicon beam and the comb-drive. $\lambda_1$ and $\lambda_2$ describe the normalized effective springs, $\mu$ is the normalized mass ratio between the comb-drive and the silicon beam, and $\varGamma$ indicates the coupling between these two oscillations. All these parameters are related to $V_{cd}$ and $V_{sg}$ via the static bending of the silicon beam, and the intrinsic properties of the device: the Young's modulus, the spring constant, the mass density, the geometry and the capacitive couplings. The extracted resonance frequencies are as follows,
\begin{align}\label{Eqthsolution1}
		\omega_1&=\sqrt{\frac{\mu\lambda_1+\lambda_2-\sqrt{4\varGamma^2\mu+(-\mu\lambda_1+\lambda_2)^2}}{2\mu}},\nonumber\\
		\omega_2&=\sqrt{\frac{\mu\lambda_1+\lambda_2+\sqrt{4\varGamma^2\mu+(-\mu\lambda_1+\lambda_2)^2}}{2\mu}}.
\end{align}

The frequencies obtained with Eq.~\ref{Eqthsolution1} are in qualitatively good agreement with experimental data (blue and red curves in Fig.~\ref{fig3}a and \ref{fig3}b), which allows us to specify the mode type to the HSM and the LSM.
According to the theory, the HSM depends on the induced tensile strain in the silicon beam with the comb-drive actuator and is therefore attributed to the silicon beam; the LSM does not show a strong dependence on $V_{cd}$ and $V_{sg}$, and represents the resonance of the comb-drive actuator. The fact that these modes are so close together, suggests that they are strongly coupled due to the clamping of the silicon beam on the comb-drive actuator. 
Note that the theory shows quantitatively good agreement with the experiment in Fig.~\ref{fig3}a. However, it does not describe the experimental SG voltage dependence very well (see Fig.~\ref{fig3}b), which is due to the following approximations in the theory: the transverse force acts on the entire length of the beam and the neglected capacitive softening.

We find an additional proof for two strongly coupled modes by comparison with a very similar device but with fixed comb-drive (see Suppl. Mat. III). In this case, we observe only a single resonant mode, indicating that the mechanical system lost 1 degree of freedom, which is the motion of the comb-drive. In addition, the quality factor in this case is $\sim$ 10 000 which is approximately 10 to 100 times larger than the quality factor of the device where the comb-drive actuator is movable (see Suppl. Mat. IV). This suggests that a significant amount of energy is transferred from the silicon beam into the comb-drive actuator and vice versa.

%\section{Coupled resonator system}
To investigate the actual effective spring and effective coupling rate, the system of two strongly coupled resonators is described by the following system of equations,
\begin{align}
(2\pi)^{-2} \omega^2 x_1 + \lambda^2  x_1 +  \gamma^2  x_2 = 0,\nonumber\\
(2\pi)^{-2} \omega^2 x_2 + \lambda^2 x_2 + \gamma^2  x_1 = 0,\label{Eq-CDE}
\end{align}
where $x_1$ and $x_2$ are effective displacements. The solutions are the two resonance frequencies $\omega_1$ and $\omega_2$ defined by Eq.~\ref{Eqthsolution1}. Note that $\lambda^2= \sfrac{1}{2}\left(\omega_1^2+\omega_2^2 \right)(2\pi)^{-2} $ is the (squared) effective spring of the system and $\gamma^2=\sfrac{1}{2}\left| \omega_1^2-\omega_2^2 \right| (2\pi)^{-2}$ is the (squared) effective coupling rate, which are listed in Table~\ref{table1} for specific $V_{cd}$ and $V_{sg}$. Surprisingly, the extracted effective spring and coupling rate show completely different behavior (see Fig.~\ref{fig3}c and \ref{fig3}d). The effective spring $\lambda$ is only weakly dependent on both $V_{cd}$ and $V_{sg}$. A maximum change in $\lambda$ of only 0.5\% is observed in our measurements, which could be attributed to the large difference in masses of the two oscillating parts. In contrast, the effective coupling rate $\gamma$ can be increased up to $10$\% by applying tensile strain through $V_{cd}$, or it can be decreased down to $10$\% by applying an out-of-plane bending through $V_{sg}$. Therefore, our device geometry gives us an independent knob to control the coupling between the two resonators.

% TABLE 1: Coupling rate
\begin{table}[t]
	\footnotesize
	\begin{tabular}{m{0.2\linewidth}|m{0.17\linewidth}<{\centering}m{0.19\linewidth}<{\centering}|m{0.17\linewidth}<{\centering}m{0.19\linewidth}<{\centering}}
		\hline
		\hline
		\multirow{2}{*}{~} & \multicolumn{2}{c|}{\rule{0pt}{0.35cm}$V_{sg}=10\text{V}$} & \multicolumn{2}{c}{\rule{0pt}{0.35cm}$V_{cd}=10\text{V}$}\\
		
		& \vspace{0.1cm} $V_{cd}=10\text{V}$ &\vspace{0.1cm} $V_{cd}=-10\text{V}$ &\vspace{0.1cm} $V_{sg}=10\text{V}$ & \vspace{0.1cm}$V_{sg}=-10\text{V}$\\
		\hline
		\vspace{0.1cm}Coupling rate [MHz] & 1.72 & 1.75 & 1.48 & 1.47\\
		
		\vspace{0.1cm}Effective spring [MHz] & 11.60 & 11.60 & 11.57 & 11.57\\
		\hline
		\hline
	\end{tabular} 
	\caption{The absolute values of the coupling rates and effective springs for different $V_{cd}$ and $V_{sg}$.}
		\label{table1}
\end{table}

\begin{figure}[t]
	\begin{center}
		\includegraphics[width=86mm]{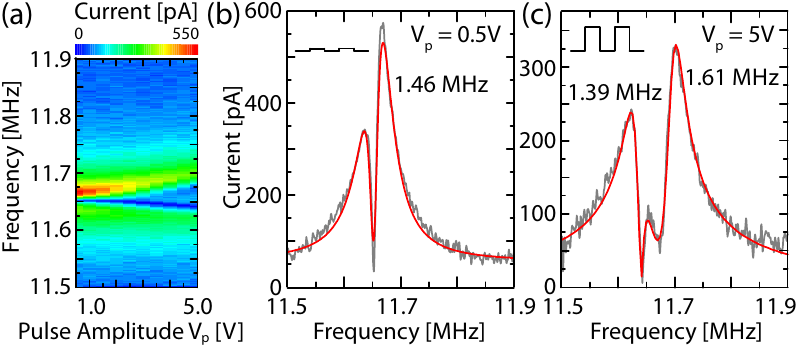}
	\end{center}
	\setlength{\abovecaptionskip}{-0.2cm}
	\caption{
		(a) The down mixed current as a function of the square wave amplitude send through the comb-drive actuator. The period of the square wave (10 kHz) is far above the response time of the observed resonances. Therefore, for large enough square wave amplitude, the resonance frequency splits into two (shown in (b) and (c) with coupling rates), showing the possibility to switch the coupling rate between the two resonators almost instantaneously.}
	\label{fig4}
\end{figure}

%\section{Pulse gating the comb-drive}
As the effective coupling rate $\gamma$ is tunable over 10\% with the comb-drive actuator, we can use $V_{cd}$ as a control voltage to switch between different coupling rates. Theoretically, it is even possible to fast switch the coupling between the two attached resonators, showing the feasibility of mechanical logic gates. To show that one can indeed use the comb-drive actuator as a switch, we applied square pulses with a frequency of 10 kHz to the comb-drive actuator (see Fig.~\ref{fig4}). The period of this frequency was chosen to be far above the response time of the comb-drive actuator and the silicon beam, which were estimated from the resonance frequency and its quality factor to be 36 $\mu$s and 12 $\mu$s respectively. For large square pulse amplitudes, the resonance frequency indeed splits into two, which shows the possibility to switch the coupling rate between the two resonators. Here, the coupling rate is switched between 1.39 MHz and 1.61 MHz at the pulse amplitude 5 V.

% FIG 4: Coupling rate tunability

%\section{Outlook and Conclusion}
In conclusion, we presented an electromechanical system, in which a silicon beam was attached to a comb-drive actuator to control the tension that operates at cryogenic temperatures. By measuring the resonance frequencies of the system, we showed that the device behaves as two strongly coupled resonators. Surprisingly, the effective coupling rate ($\sim 1.5$ MHz) is tunable with the comb-drive actuator (+10\%) as well as with a side-gate (-10\%) placed close to the silicon beam. In contrast, the effective spring constant of the system is insensitive to an applied voltage to either the comb-drive actuator or the side-gate. From an applied point of view, the high tunability of coupling rate would be very interesting if two resonators are attached to the comb-drive actuator as multifunctional force sensors.  Finally, we showed that the comb-drive actuator can be used to switch between different coupling rates with a frequency of at least 10 kHz, which is particularly interesting for future mechanical logic devices in which it is necessary to fast switch between two states.

\section{Acknowledgement}

We acknowledge support from the Helmholtz Nanoelectronic Facility (HNF) and funding from the ERC (GA-Nr. 280140).

\end{document}